\documentclass[openacc]{rstransa}
\usepackage{graphicx}
\usepackage{epsfig}
\usepackage[caption=false]{subfig}

\begin{document}

\title{Analogue Gravity on a Superconducting Chip}

\author{Miles P. Blencowe and Hui Wang}

\address{Department of Physics and Astronomy, Dartmouth College, Hanover, New Hampshire 03755, USA}

\subject{quantum physics, particle physics, optics}


\corres{Miles Blencowe\\
\email{blencowe@dartmouth.edu}}

\begin{abstract}
We describe how analogues of a Hawking  evaporating black hole as well as the Unruh effect for an oscillatory, accelerating photodetector in vacuum may be realized using superconducting, microwave circuits that are fashioned out of Josephson tunnel junction and film bulk acoustic resonator elements.   
\end{abstract}

\maketitle

\section{Introduction}
In an earlier publication\cite{nation2009}, we proposed a possible way to demonstrate a Hawking radiation analogue by utilizing a superconducting microwave transmission line comprising an array of direct-current superconducting quantum interference devices (dc-SQUID's). This proposal was directly inspired by  related analogue Hawking effect schemes involving electromagnetic waveguides \cite{schutzhold2005} and  non-linear optical fibres \cite{philbin2008}. One advantage to utilizing superconducting capacitor and inductor elements for the waveguide is that they are low noise, with quantum zero-point fluctuations dominating over thermal radiation fluctuations at standard, operating temperatures of tens of milliKelvins and below for ultrahigh vacuum-dilution fridge set-ups. Furthermore, with Josephson junctions functioning effectively as strongly non-linear inductors, the quantum fluctuations can be converted into real microwave photons analogously to the Hawking and dynamical Casimir effects \cite{wilson2011,lahteen2013,nation2012}. Finally, by incorporating also micrometre scale, mechanically vibrating crystals in the gigahertz frequency range, close analogues of both the dynamical Casimir \cite{sanz2018}  and Unruh effects \cite{wang2019} may be realized.

Since the publication of our  proposal\cite{nation2009}, there have been several related theoretical investigations concerning analogue gravity vacuum photon production processes that utilize superconducting transmission lines. For example, in Ref. \cite{tian2017}, a superconducting circuit analogue of cosmological particle creation was proposed (see also Ref. \cite{lang2019}). 
There have also been significant advances in superconducting microwave circuit technology, largely driven by the push to realize quantum computers implementing superconducting qubits; in particular, the recently developed  Josephson traveling-wave parametric amplifier (JTWPA) \cite{macklin2015} is ideally suited (with some modifications) for demonstrating the analogue Hawking effect with detectable microwave photon production.

In this paper, we begin with a description of our analogue Hawking effect scheme in light of recent superconducting circuit technology developments. We also clarify and improve some aspects of our original analysis in Ref. \cite{nation2009}. In the second part, we outline a possible scheme for realizing a close analogue of the Unruh effect for an oscillating (actual accelerating) superconducting qubit photodetector by utilizing a so-called film bulk acoustic resonator (FBAR) element in addition to capacitor and Josephson junction elements. The following discussions will emphasize near-future, possible directions for theoretical and experimental exploration.  

\section{Analogue Hawking Effect}
\begin{figure}[!htb]
\centering\includegraphics[width=4in]{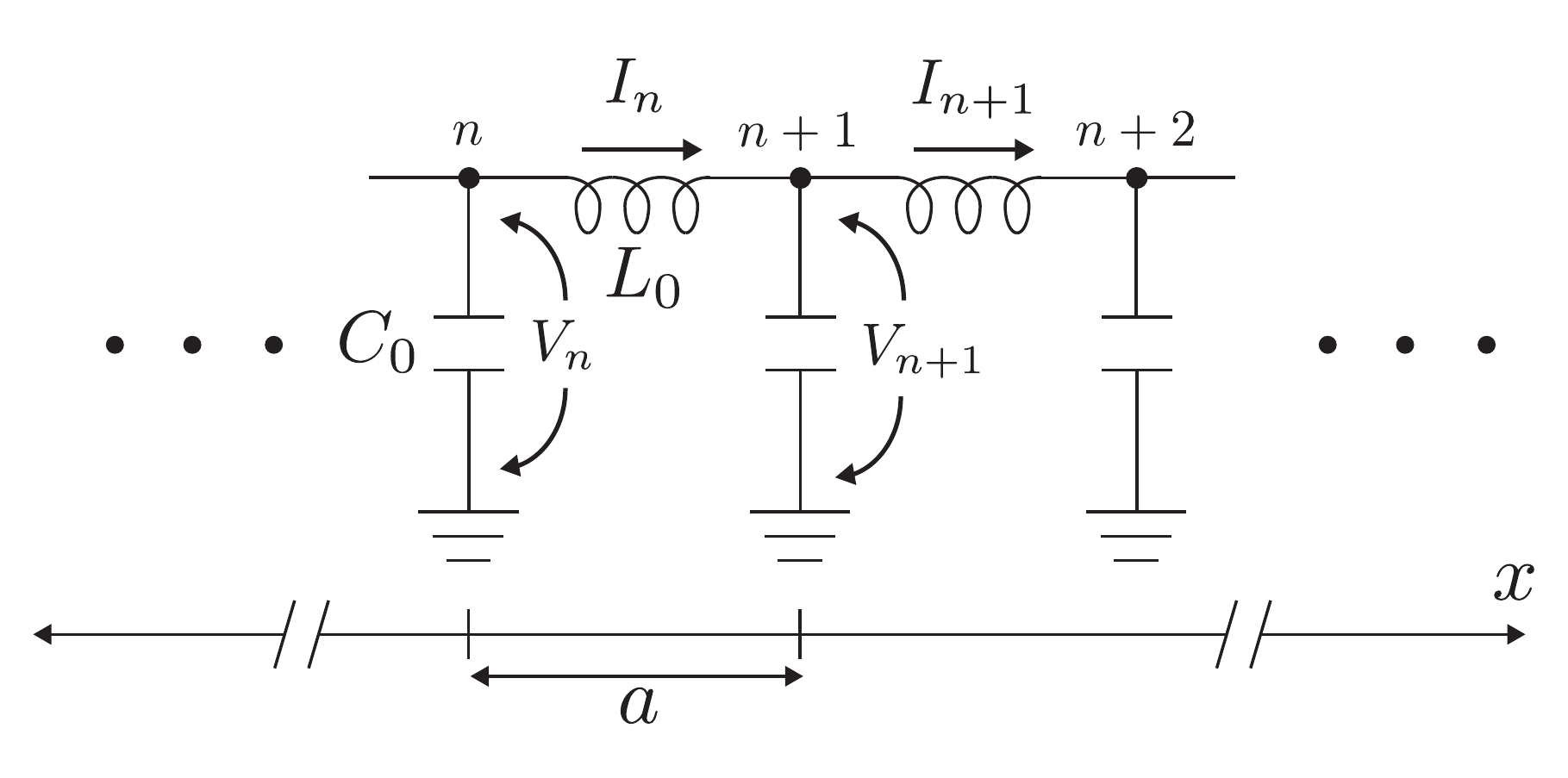}
\caption{Circuit diagram representation of a microwave transmission line comprising inductors in series and capacitors in parallel, where each connected inductor ($L_0$) and capacitor ($C_0$) element forms a unit cell of length $a$.}
\label{transfig}
\end{figure}

We begin our discussion by considering a simple, infinitely long one-dimensional transmission line comprising lumped element inductors in series and capacitors in parallel (Fig.\ref{transfig}). The inductors have inductance value $L_0$ and capacitors have capacitance value $C_0$. The transmission line is made out of repeating unit cells of length $a$. 
Applying Kirchhoff's laws, which state that the current entering a circuit node must equal the current exiting the node (current conservation) and that the sum over all voltages around a closed circuit loop must equal the time rate of change of the magnetic flux threading the loop (Faraday's law of induction) \cite{vool2017}, we have
\begin{eqnarray}
I_n-I_{n+1}&=& \frac{dQ_{n+1}}{dt},\label{kircheq}\\
V_n-V_{n+1}&=&\frac{d\Phi_n}{dt},\label{kirch2eq}
\end{eqnarray}   
where $Q_{n+1}$ is the charge on the capacitor attached to node $n+1$ and $\Phi_n$ is the magnetic flux threading the inductor linking nodes $n$ and $n+1$. Utilizing the constitutive relations between voltage and charge for the capacitor, $V_n=Q_n/C_0$, and between current and flux for the inductor, $I_n=\Phi_n/L_0$, Eqs. (\ref{kircheq}) and (\ref{kirch2eq}) become
 \begin{eqnarray}
I_n-I_{n+1}&=& C_0\frac{dV_{n+1}}{dt},\label{kirch3eq}\\
V_n-V_{n+1}&=&L_0\frac{dI_{n}}{dt}\label{kirch4eq}.
\end{eqnarray}
We now introduce a generalized, dimensionless flux coordinate $\varphi_n$ associated with the capacitor voltages \cite{vool2017}:
\begin{equation}
   V_n=\frac{\Phi_0}{2\pi}\frac{d\varphi_n}{dt},
    \label{genflux}
\end{equation}
where $\Phi_0=h/(2e)$ is the magnetic flux quantum. From now on, will call `$\varphi_n$' the `phase' coordinate; the advantages to working with this phase coordinate will become apparent when we introduce the Josephson junction element below. Substituting Eq. (\ref{genflux}) into Eq. (\ref{kirch4eq}), we obtain the following expression for the current in terms of the phase coordinate:
\begin{equation}
    I_n=-\frac{\Phi_0}{2\pi L_0}\left(\varphi_{n+1}-\varphi_n\right),
    \label{currentphaseeq}
\end{equation}
and further substituting Eqs. (\ref{genflux}) and (\ref{currentphaseeq}) into Eq. (\ref{kirch3eq}) gives the transmission line equation of motion in terms of the phase coordinate
\begin{equation}
    \frac{d^2\varphi_n}{dt^2}=\frac{1}{L_0 C_0}\left(\varphi_{n+1}-2\varphi_n+\varphi_{n-1}\right).
    \label{phaseeomeq}
\end{equation}
These equations can be obtained from the Lagrangian
\begin{equation}
    L=\sum^{+\infty}_{-\infty}\left[\frac{1}{2} C_0\left(\frac{\Phi_0}{2\pi}\right)^2 \left(\frac{d\varphi_n}{dt}\right)^2-\frac{1}{2L_0}\left(\frac{\Phi_0}{2\pi}\right)^2 \left(\varphi_{n+1}-\varphi_n\right)^2\right]
    \label{phaselageq}
\end{equation}
via Lagrange's equations, and from the associated Hamiltonian
\begin{equation}
H=\sum^{+\infty}_{n=-\infty} \left[\left(\frac{2\pi}{\Phi_0}\right)^2\frac{p_n^2}{2C_0}+\left(\frac{\Phi_0}{2\pi}\right)^2\frac{\left(\varphi_{n+1}-\varphi_n\right)^2}{2L_0}\right]
\label{hameq}
\end{equation}
via Hamilton's equations, where the momentum conjugate to the phase  coordinate $\varphi_n$ is $p_n=\left(\Phi_0/2\pi\right)^2 C_0d\varphi_n/dt=\left(\Phi_0/2\pi\right) C_0V_n= \left(\Phi_0/2\pi\right) Q_n$.

From Eqs. (\ref{genflux})-(\ref{hameq}), we see that the transmission line supports coupled, propagating current and voltage waves. In particular, Eq. (\ref{phaseeomeq}) has propagating eigenfrequency solutions with dispersion relation
\begin{equation}
\omega(k)=\frac{2}{\sqrt{L_0C_0}}\left|\sin\left(\frac{k a}{2}\right)\right|,
\label{dispeq}
\end{equation}
giving an upper cut-off frequency $\omega_c=2/\sqrt{L_0C_0}$, corresponding to a minimum wavelength $\lambda_c=a$.
Restricting to long wavelength solutions $\lambda\gg a$, the dispersion relation (\ref{dispeq}) becomes approximately linear: $\omega=c|k|$, where $c=1/\sqrt{{\mathcal{LC}}}$ is the electromagnetic wave speed along the transmission line and ${\mathcal{C}}=C_0/a$ and ${\mathcal{L}}=L_0/a$ denote the capacitance and inductance per unit length respectively of the transmission line. In this long wavelength, continuum limit, we can approximate the current and voltage differences in Eqs. (\ref{kirch3eq}) and (\ref{kirch4eq}) as $(I_{n+1}-I_{n})/a \approx \partial I/\partial x$ and $(V_{n+1}-V_{n})/a \approx \partial V/\partial x$ respectively, to obtain
\begin{eqnarray}
\frac{\partial I(x,t)}{\partial x}&=& -{\mathcal{C}}\frac{\partial V(x,t)}{\partial t},\label{kirch5eq}\\
\frac{\partial V(x,t)}{\partial x}&=& -{\mathcal{L}}\frac{\partial I(x,t)}{\partial t}\label{kirch6eq},
\end{eqnarray}
where . The continuum limits of Eqs. (\ref{genflux}) and (\ref{currentphaseeq}) expressing the voltage and current respectively in terms of the phase field coordinate $\varphi(x,t)$ are
\begin{equation}
   V(x,t)=\frac{\Phi_0}{2\pi}\frac{\partial\varphi (x,t)}{\partial t},
    \label{genflux2}
\end{equation}
\begin{equation}
    I(x,t)=-\frac{\Phi_0}{2\pi {\mathcal{L}}}\frac{\partial\varphi(x,t)}{\partial x}.
    \label{currentphase2eq}
\end{equation}
Submitting these expressions in Eq. (\ref{kirch5eq}), we obtain the continuum limit to the equation of motion (\ref{phaseeomeq}):
\begin{equation}
    \frac{\partial^2\varphi}{\partial t^2}=c^2 \frac{\partial^2\varphi}{\partial x^2},
    \label{waveeq}
\end{equation}
which follows via Hamilton's equations from the continuum limit of Hamiltonian (\ref{hameq}):
\begin{equation}
H=\int_{-\infty}^{+\infty} dx\left[\left(\frac{2\pi}{\Phi_0}\right)^2\frac{p(x,t)^2}{2{\mathcal{C}}}+\left(\frac{\Phi_0}{2\pi}\right)^2\frac{\partial_x\varphi(x,t)^2}{2{\mathcal{L}}}\right].
\label{conthameq}
\end{equation}
where  $p(x,t)=\left(\Phi_0/2\pi\right)^2{\mathcal{C}}\partial_t \varphi(x,t)=\left(\Phi_0/2\pi\right){\mathcal{C}}V(x,t)$ is the momentum conjugate to the phase coordinate $\varphi(x,t)$. 
 Equation (\ref{waveeq}) is the wave equation for a massless scalar field. This equation can be written as an effective, relativistic Klein Gordon equation for a massless scalar field   in $1+1$ spacetime dimensions:
 \begin{equation}
 \partial_{\mu}\partial^{\mu}\varphi=0,
 \label{KGeq}
 \end{equation}
 where $\partial_{\mu}=(\partial/\partial t,\partial/\partial x)$ and  we have introduced an effective  Minkowski spacetime metric:
\begin{equation}
\eta^{\mu\nu}=
\begin{pmatrix} 
-1 & 0 \\
0&c^{2}
\end{pmatrix}. 
\label{paulirleq}
\end{equation}

With the transmission line made from a common superconducting element such as aluminium or niobium, cooling below the critical temperature results in low loss-low noise electromagnetic wave propagation in the microwave frequency regime; quantum behaviour such as vacuum fluctuations can become apparent and hence the transmission line must be modeled as a quantum system. The equal time canonical commutation relation  for the continuum field operator and its conjugate momentum operator in the Heisenberg picture is:
\begin{equation}
\left[\hat{\varphi}(x,t),\hat{p}(x',t)\right]=i\hbar\delta(x-x'),
\label{ccreq}
\end{equation} 
with all other commutation relations vanishing, and where recall $\hat{p}(x,t)={\mathcal{C}}(\Phi_0/2\pi)^2 \partial_t \hat{\varphi}(x,t)$. The field operator equations of motion in the Heisenberg picture are $ \partial_{\mu}\partial^{\mu}\hat{\varphi}=0$ and can be solved in terms of photon creation and annihilation operators as
\begin{equation}
\hat{\varphi}(x,t)=\frac{2\pi}{\Phi_0}\sqrt{\frac{\hbar Z}{4\pi}}\int_{-\infty}^{+\infty}\frac{dk}{\sqrt{|k|}}\left[e^{-i\left(\omega t-kx\right)}\hat{a}(k) +e^{+i\left(\omega t-kx\right)}\hat{a}^{\dag}(k)\right],
\label{aeq}
\end{equation}
with linear dispersion relation $\omega=c|k|$ and where $Z=\sqrt{{\mathcal{L}}/{\mathcal{C}}}$ is the transmission line impedance.  From Eqs.  (\ref{ccreq}) and (\ref{aeq}), the creation and annihilation operators satisfy the commutation relations
\begin{equation}
\left[\hat{a}(k),\hat{a}^{\dag}(k')\right]=\delta(k-k').
\label{acreq}
\end{equation} 
Photon states are created from the vacuum defined through $\hat{a}(k,0)\left|0\right.\rangle=0$, with single photon states $\hat{a}^{\dag}(k)\left|0\right.\rangle$ having energy $E=\hbar\omega$ and momentum $p=\hbar k$. This quantum field model describes photons propagating in an effective $1+1$ dimensional Minkowski spacetime. Such a continuum approximation is a good one provided photon energies are much smaller than the `Planck' energy $hc/a$ set by the transmission line unit cell size $a$. 

Note that Eq. (\ref{aeq}) is infrared ($k\rightarrow 0$) divergent, a consequence of the photons being massless. However, the corresponding expressions for measurable quantities such as the voltage (\ref{genflux2}) or current (\ref{currentphase2eq}) are not infrared divergent. Furthermore, current and voltage measurements are typically filtered through some bandwidth $\Delta\omega$ centred at some frequency $\omega_0$ of interest in the microwave (i.e., GHz) range, where $\Delta\omega\ll\omega_0$. Provided $\omega_c\gg\omega_0$, or equivalently $a\ll\lambda_0$, then the ultraviolet (i.e., upper cutoff $\omega\rightarrow\omega_c$) corrections to Eq. (\ref{aeq}) will not affect the measurement results.        

Is it possible however to modify the superconducting transmission line circuit in some way so as to recover a continuum effective  quantum field theory on a spacetime with an event horizon? The answer to this question is indeed `yes'! We require a distinct type of circuit element called a Josephson tunnel junction (JJ)\cite{vool2017}, which comprises two overlapping, superconducting metal electrodes separated by an electrically insulating, nanometre thick oxide layer, such that  Cooper pairs--the superconducting charge carriers--are able to quantum tunnel between the electrodes. A lumped element model of a JJ satisfies the following respective relations for the voltage drop across the junction and current flow through the junction:
\begin{equation}
V_J=\frac{\Phi_0}{2\pi} \frac{d\varphi_J}{dt},
\label{vjeq}
\end{equation}
\begin{equation}
I_J=I_c \sin\varphi_J,
\label{ijeq}
\end{equation}
where  the coordinate $\varphi_J$ is the gauge invariant phase difference across the tunnel junction of the macroscopic wave function describing the Cooper pairs and $I_c$ is the critical current of the tunnel junction, i.e., the maximum possible dc Cooper-pair tunnel current. Comparing the voltage expression with that for an inductor [see, Eq. (\ref{kirch2eq})], we can define an effective flux variable for the Josephson junction:  $\Phi_J=\frac{\Phi_0}{2\pi}\varphi_J$. For small phase magnitudes $|\varphi|({\mathrm{mod}}\, 2\pi n)\ll 1$, Eq. (\ref{ijeq}) can be approximately expressed as $I_J\approx \frac{2\pi I_c}{\Phi_0} \Phi_J$, and comparing with the constitutive relation given above between current and flux for the inductor, we see that the JJ functions as an effective inductor with inductance $L_J\approx \frac{\Phi_0}{2\pi I_c}$. For not necessarily small phase magnitudes, we can interpret the JJ as an effective phase coordinate-dependent inductor:
\begin{equation}
L_J(\varphi_J)=\frac{\Phi_J}{I_c\sin(2\pi \Phi_J/\Phi_0)}=\frac{\Phi_0\varphi_J}{2\pi I_c \sin\varphi_J}.
\label{jinducteq}
\end{equation}
It is this phase/flux dependent effective inductance of the JJ  that will enable the realization of an analogue $1+1$ dimensional spacetime with event horizon as we now show. 

\begin{figure}[!htb]
\centering\includegraphics[width=4.5in]{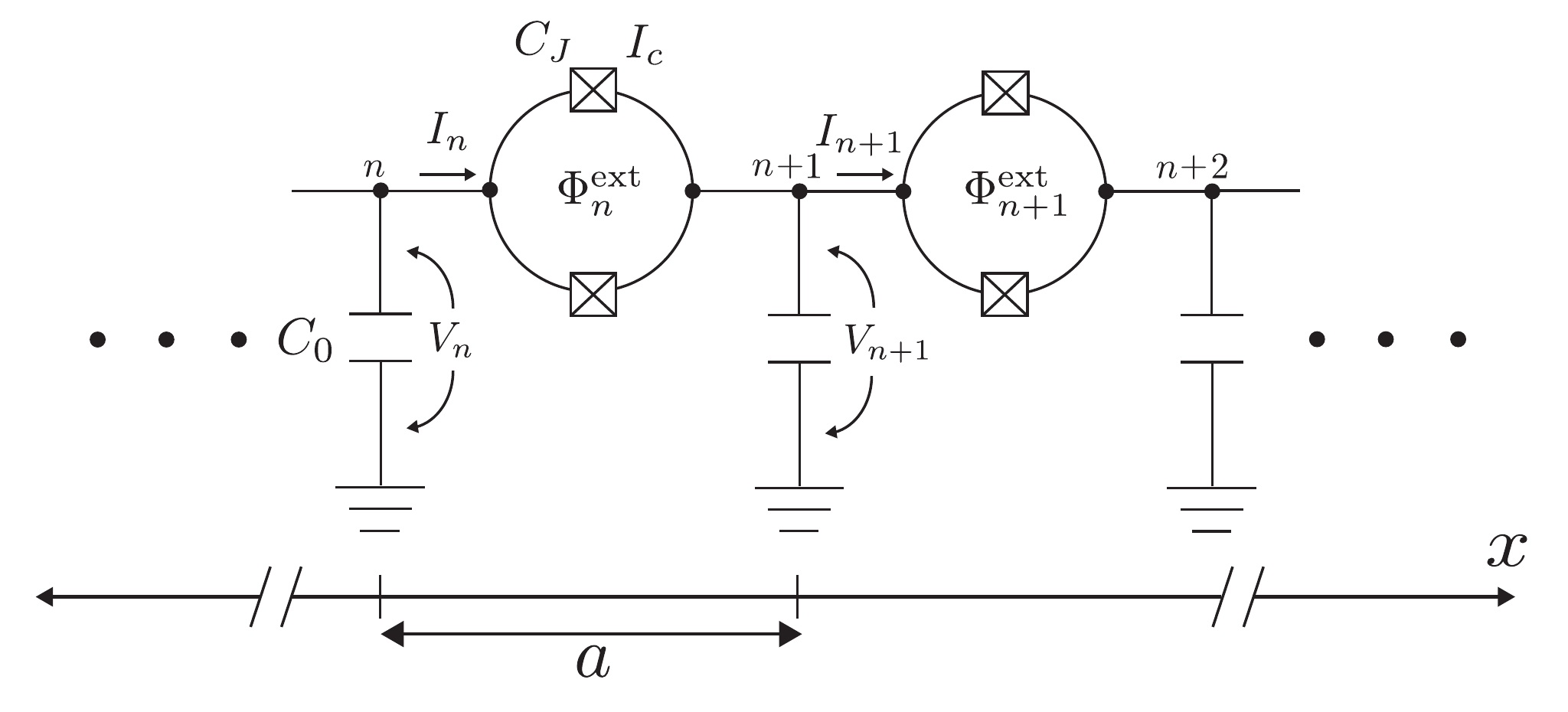}
\caption{Circuit diagram of an array of dc-SQUID's  in series and capacitors in parallel, where each connected dc-SQUID and capacitor ($C_0$) element forms a unit cell of length $a$. The dc-SQUID's are biased with an external magnetic flux that is allowed to vary with both position (hence the $n$ label) and time.}
\label{JJtransfig}
\end{figure}

In particular, we replace each inductor element in the transmission line circuit of Fig. \ref{transfig} with two JJ's in parallel--called a dc-SQUID (Fig. \ref{JJtransfig}).
The JJ elements (symbolized by the crossed boxes)  making up the dc-SQUID's are assumed to have identical critical current values $I_c$ and capacitance values $C_J$. The dc-SQUID's are threaded with an external magnetic flux that is allowed to vary from unit cell to unit cell, as well as depend on time. The dc-SQUID array transmission line is an example of a `metamaterial', which enables microwave propagation dynamics that would not be possible in a simple co-planar transmission line fashioned from a centre superconductor strip and ground plane. The circuit shown in Fig. \ref{JJtransfig} is also closely related to Josephson junction traveling-wave parametric amplifiers (JTWPA's), which can now be realised with very small deviations in the repeating unit cell JJ and capacitor values \cite{macklin2015,zorin2019}. Not shown is the circuitry necessary for providing the space-time dependent flux bias, which would involve another, current biased transmission line metamaterial running in parallel to the indicated dc-SQUID array \cite{zorin2019}.

Applying Kirchhoff's laws to the circuit shown in Fig. \ref{JJtransfig}, we have [c.f. Eqs. (\ref{kircheq}) and (\ref{kirch2eq})]:
\begin{eqnarray}
I_n-I_{n+1}&=& \frac{dQ_{n+1}}{dt},\label{JJkircheq}\\
V_n-V_{n+1}&=&\frac{\Phi_0}{2\pi}\frac{d\varphi_{Jn}}{dt},\label{JJkirch2eq},
\end{eqnarray}
where we have used Eq. (\ref{vjeq}) with $\varphi_{Jn}$ the average of the sum of the phases across each of the two JJ's forming the dc-SQUID loop in the $n$th unit cell. The current entering the $n$th unit cell dc-SQUID expressed in terms of the JJ average phase is \begin{equation}
    I_n=2C_J \frac{\Phi_0}{2\pi} \frac{d^2\varphi_{J n}}{dt^2}+2 I_c \cos\left(\frac{\pi\Phi^{\mathrm{ext}}_n}{\Phi_0}\right) \sin\varphi_{J n},
    \label{squidieq}
\end{equation}
while voltage across each capacitor is related to the charge as $V_n=Q_n/C_0$. Similarly to the above analysis for the inductor-capacitor transmission line, it is more convenient to express the dynamics in terms of the phase coordinates $\varphi_n$ across the $C_0$ capacitors instead of the JJ phase coordinates $\varphi_{Jn}$, where from Eq. (\ref{JJkirch2eq}) we have  $\varphi_{Jn}=\varphi_n-\varphi_{n+1}$ (up to an inessential constant),  with $\dot{\varphi}_n=2 \pi V_n/\Phi_0$; in particular, the system Lagrangian is more straightforward to write down. Substituting Eq. (\ref{squidieq}) into  Eq. (\ref{JJkircheq}) gives
for the dc-SQUID array equations of motion:
\begin{eqnarray}
      &&\left(C_0+4C_J\right)\left(\frac{\Phi_0}{2\pi}\right)^2 \frac{d^2\varphi_n}{dt^2}-2 C_J \left(\frac{\Phi_0}{2\pi}\right)^2\left(\frac{d^2\varphi_{n-1}}{dt^2}+\frac{d^2\varphi_{n+1}}{dt^2}\right)\cr
      &&=-2E_J \cos\left(\pi\Phi^{\mathrm{ext}}_n/\Phi_0\right) \sin\left(\varphi_n-\varphi_{n+1}\right)+2E_J \cos\left(\frac{\pi\Phi^{\mathrm{ext}}_{n-1}}{\Phi_0}\right) \sin\left(\varphi_{n-1}-\varphi_n\right),\cr
      &&\label{JJeomeq}
\end{eqnarray}
where $E_J=\Phi_0 I_c/(2\pi)$ is the so-called `Josephson energy'. Equation (\ref{JJeomeq}) follows via Lagrange's equations from the Lagrangian
\begin{eqnarray}
&&L=\sum^{+\infty}_{n=-\infty}\bigg[\frac{1}{2}C_0 \left(\frac{\Phi_0}{2\pi}\right)^2 \left(\frac{d\varphi_n}{dt}\right)^2+C_J \left(\frac{\Phi_0}{2\pi}\right)^2\left(\frac{d\varphi_n}{dt}-\frac{d\varphi_{n+1}}{dt}\right)^2\cr 
&&+2E_J \cos\left(\frac{\pi\Phi^{\mathrm{ext}}_{n}}{\Phi_0}\right) \cos\left(\varphi_{n}-\varphi_{n+1}\right)\bigg],
\label{squidarrayeq}
\end{eqnarray}
where we assume for simplicity an ideal, infinitely long dc-SQUID array; an actual dc-SQUID array will comprise several thousand unit cells \cite{macklin2015}.

Considering long wavelength $\lambda\gg a$ dynamics, where $\varphi_n$ changes little from one unit cell to the next, we can make the approximation $(\varphi_{n+1}-\varphi_n)/a\approx \partial\varphi/\partial x$.  The resulting continuum approximation for the Lagrangian (\ref{squidarrayeq}) is 
\begin{eqnarray}
&L=\int_{-\infty}^{+\infty}dx \bigg[\frac{1}{2}{\mathcal{C}}\left(\frac{\Phi_0}{2\pi}\right)^2\left(\frac{\partial\varphi}{\partial t}\right)^2 +C_J a \left(\frac{\Phi_0}{2\pi}\right)^2 \left(\frac{\partial^2\varphi}{\partial t\partial x}\right)^2
-E_J a  \cos\left(\frac{\pi\Phi^{\mathrm{ext}}(x,t)}{\Phi_0}\right) \left(\frac{\partial\varphi}{\partial x}\right)^2\biggr],\cr
&\label{contjjlageq}
\end{eqnarray}
where ${\mathcal{C}}=C_0/a$. We can furthermore neglect the second term in the Lagrangian provided the JJ capacitances satisfy $C_J\ll C_0 (\lambda/a)^2$, and the resulting approximate Hamiltonian is 
\begin{equation}
    H=\int_{-\infty}^{+\infty}dx\left[\left(\frac{2\pi}{\Phi_0}\right)^2 \frac{\left(p(x,t)\right)^2}{2{\mathcal{C}}}+E_J a  \cos\left(\frac{\pi\Phi^{\mathrm{ext}}(x,t)}{\Phi_0}\right) \left(\frac{\partial\varphi}{\partial x}\right)^2\right],
    \label{jjarrayhameq}
\end{equation}
where the momentum conjugate to the phase field coordinate $\varphi(x,t)$ is $p(x,t)={\mathcal{C}}(\frac{\Phi_0}{2\pi})^2\partial_t\varphi(x,t)$. Hamilton's equations give the following wave equation
\begin{equation}
    -\frac{\partial^2\varphi}{\partial t^2}+\frac{\partial}{\partial x}\left(c^2(x,t) \frac{\partial\varphi}{\partial x}\right)=0,
    \label{jjwaveeq}
\end{equation}
where the spacetime dependent electromagnetic wave phase speed is
\begin{equation}
    c(x,t)=\frac{1}{\sqrt{{\mathcal{L}}(x,t){\mathcal{C}}}},
    \label{JJceq}
\end{equation}
with the effective inductance per unit length given by
\begin{equation}
    {\mathcal{L}}(x,t)=\frac{\Phi_0}{4\pi I_c a} \sec\left(\frac{\pi\Phi^{\mathrm{ext}}(x,t)}{\Phi_0}\right).
    \label{effindeq}
\end{equation}
The phase speed inherits its spacetime dependence from the varying magnetic flux threading the dc-SQUID loops of the array via the effective inductance of the dc-SQUIDs. The wave equation (\ref{jjwaveeq}) can also be directly obtained by taking the continuum limit of the dc-SQUID array equations (\ref{JJeomeq}).

\begin{figure}[!htb]
\centering\includegraphics[width=3in]{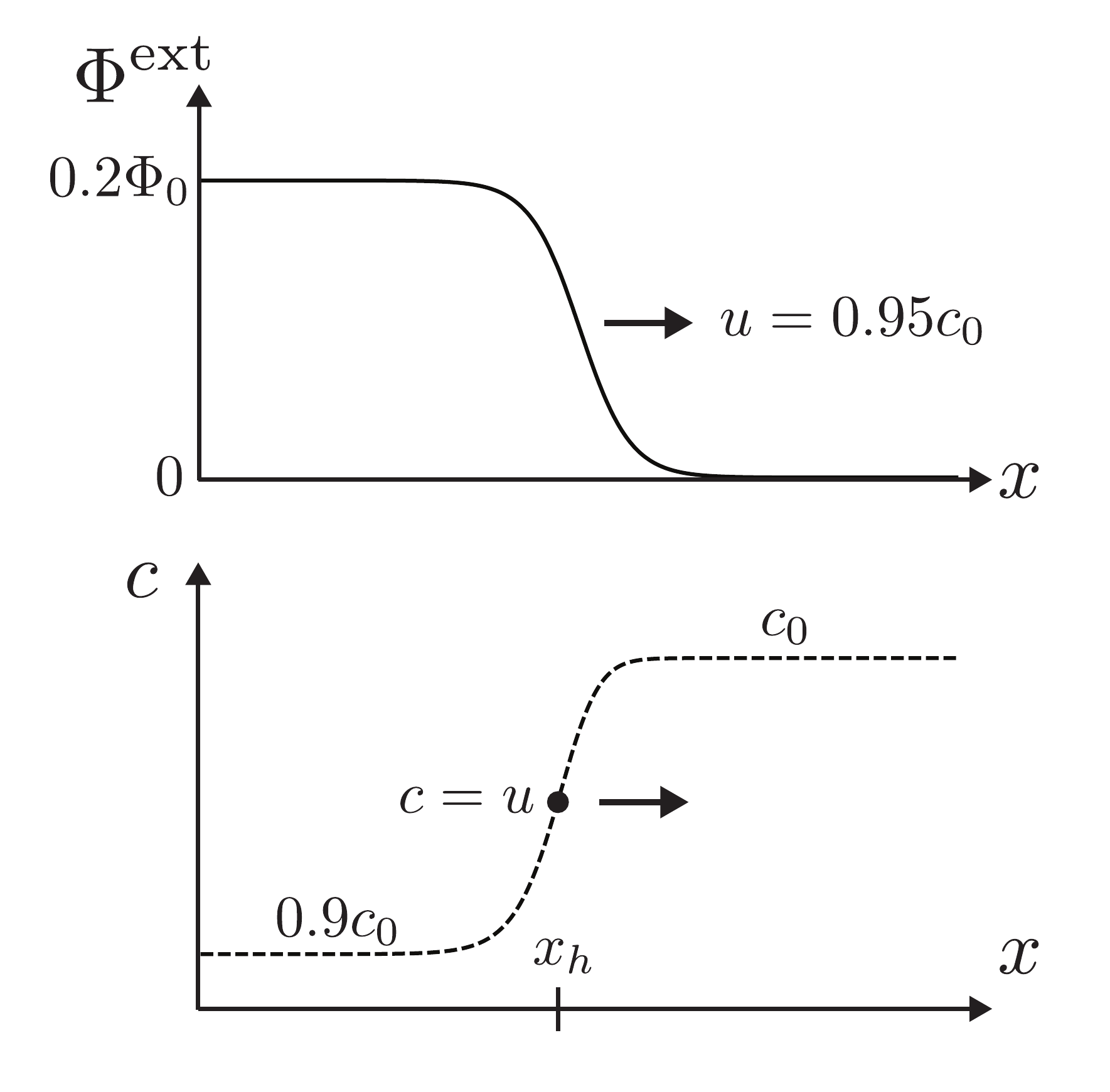}
\caption{Propagating external flux bias  step with front speed $u=0.95 c_0$ (top plot) and the resulting electromagnetic wave phase speed in the dc-SQUID array (bottom plot). The marked location $x_h(t)$ on the phase speed plot where $c(x_h,t)=u=0.95 c_0$ constitutes an effective moving event horizon.}
\label{pulsefig}
\end{figure}
Notice from Eqs. (\ref{JJceq}) and (\ref{effindeq}) that increasing the external flux bias $\Phi^{\mathrm{ext}}$ from zero increases the effective inductance and hence decreases the electromagnetic wave phase speed. This is made more transparent by combining Eqs. (\ref{effindeq}) and (\ref{JJceq}) in the form
\begin{equation}
  c(x,t)=c_0 \sqrt{\cos\left(\frac{\pi\Phi^{\mathrm{ext}}(x,t)}{\Phi_0}\right)},
  \label{speedeq}
\end{equation}
where $c_0$ is the phase speed that follows from Eqs. (\ref{JJceq}) and (\ref{effindeq}) with zero external flux bias ($\Phi^{\mathrm{ext}}=0$):
\begin{equation}
    c_0={a}\sqrt{\frac{4\pi I_c}{\Phi_0 C_0}}.
    \label{zerofluxspeedeq}
\end{equation}
The ability to utilize an external flux bias to lower the phase speed suggests a way to make an effective event horizon \cite{schutzhold2005,philbin2008}. In particular, consider as an example a flux step of magnitude $\Phi^{\mathrm{ext}}=0.2\Phi_0$ with front that moves with speed $u=0.95 c_0$, and where the flux bias is zero ahead of the propagating front (Fig. \ref{pulsefig}--top plot). Substituting the flux step function into Eq. (\ref{speedeq}) gives the resulting electromagnetic phase speed function shown in the bottom plot of Fig. \ref{pulsefig}. Well to the left of the location $x_h(t)$ for which $c(x_h,t)=u=0.95 c_0$, electromagnetic waves travel with speed $0.9c_0<u$, while well to the right of the location $x_h(t)$, electromagnetic waves travel with speed $c_0>u$. Thus the location $x_h(t)$ constitutes an effective moving event horizon.

Transforming to the comoving frame of the propagating flux front, $x'=x-ut$, $t'=t$, the wave equation (\ref{jjwaveeq}) becomes 
\begin{equation}
    \left(-\frac{\partial^2}{\partial t^2}+2 u\frac{\partial^2}{\partial x\partial t}+\frac{\partial}{\partial x}\left(c^2(x)-u^2\right) \frac{\partial}{\partial x}\right)\varphi=0,
    \label{comoveq}
\end{equation}
where we have dropped the primes on the comoving coordinates. Equation (\ref{comoveq}) can be expressed  in the `general covariant' form
\begin{equation}
    \frac{1}{\sqrt{-g}}\partial_{\mu}\left(\sqrt{-g}g^{\mu\nu}\partial_{\nu}{\varphi}\right)=0,
    \label{covwaveeq}
\end{equation}
with $g={\mathrm{det}}g_{\mu\nu}$ and the effective spacetime metric defined as
\begin{equation}
   g^{\mu\nu}=
\frac{1}{c(x)}\begin{pmatrix} 
-1 & u \\
u&c^{2}(x)-u^2
\end{pmatrix}.
\label{effmeteq}
\end{equation}
This effective metric displays an event horizon where $c(x)=u$. 

Note that, while we have described the co-planar transmission line with an effective, $1+1$ dimensional wave equation for which the metric is conformally flat, as an actual device the transmission line is of course three dimensional with transverse $y$-$z$ dimensions. If we were to take into account these transverse dimensions, then we could instead write down a $1+3$ dimensional wave equation with effective metric that is not conformally flat, and hence the transmission line device with propagating flux bias step does provide an analogue of a genuine curved spacetime with event horizon \cite{schutzhold2005}. Nevertheless, with the electromagnetic wave dynamics of interest having characteristic wavelengths $\lambda$ much larger than these transverse dimensions, the above $1+1$ dimensional model should yield an accurate description of the wave dynamics.  

Quantizing the continuum model Hamiltonian (\ref{jjarrayhameq}), the equal time canonical commutation relation  for the continuum field operator and its conjugate momentum operator in the Heisenberg picture is:
\begin{equation}
\left[\hat{\varphi}(x,t),\hat{p}(x',t)\right]=i\hbar\delta(x-x'),
\label{ccr2eq}
\end{equation}
with all other commutation relations vanishing, and 
where recall $\hat{p}(x,t)={\mathcal{C}}(\Phi_0/2\pi)^2 \partial_t \hat{\varphi}(x,t)$.  The field operator equations of motion in the Heisenberg picture coincide with the classical equations (\ref{jjwaveeq}), with $\varphi\rightarrow\hat{\varphi}$.   Assuming a propagating external flux bias step as described above, then photon pair production from the electromagnetic vacuum should occur in the vicinity of the event horizon, in direct analogy to Hawking radiation; In particular, the above commutation relations are consistent with allowing the mixing of positive and negative frequencies and hence photon production from the vacuum \cite{unruh2003}. The effective Hawking temperature is \cite{schutzhold2005}
\begin{equation}
    T_H=\frac{\hbar}{2\pi k_B}\left|\frac{\partial c(x)}{\partial x}\right|_{x_h}.
    \label{hawkingeq}
\end{equation}

Let us now estimate possible Hawking temperatures that can be realized. Assuming the phase speed step  length to be about ten times the unit cell length $a$ and the step height to be $0.1 c_0$ as considered above, we have $|\partial c/\partial x|_{x_h} \approx 0.01 c_0/a$. Substituting in Eq. (\ref{zerofluxspeedeq}), we obtain
\begin{equation}
    T_H \approx \frac{1}{100\pi k_B} \sqrt{\frac{\hbar e I_c}{C_0}},
    \label{hawkingesteq}
\end{equation}
where recall that $C_0$ is the unit cell lumped element capacitance to ground (see Fig. \ref{JJtransfig}) [not to be confused with the zero bias flux phase speed $c_0$]. For the realizable example circuit values $I_c=5\, \mu{\mathrm{A}}$, $C_0 =1\, {\mathrm{fF}}$ \cite{macklin2015},
we have $T_h\approx 70\, {\mathrm{mK}}$. 

An experimental run would require the repeated launching of propagating flux bias steps down the dc-SQUID array, with the  measured Hawking emitted photon pairs' correlation signal recovered through sufficiently long time-averaging along the lines of the acoustic Hawking radiation counterpart Bose-Einstein condensate experiment of Ref.  \cite{nova2019}. In particular, for the above considered example parameter values, the emitted photon pairs propagate at speeds $0.9c_0$ and $c_0$ to the left and right of the moving event horizon, respectively, resulting in a spatial separation of the photon pairs and hence a delay in their arrival time at the photodetectors  located at the right, terminal end of the dc-SQUID array. Verification of such Hawking process generated photon pairs may be achieved through measurements of the second order correlation of the photon number \cite{balbinot2008} as well as their entanglement \cite{vidal2002,jacquet2020}. Detection methods might include linear measurements of the microwave field quadratures \cite{wilson2011,lahteen2013,eichler2012,lahteen2016,sandbo2018}, or single-shot microwave photon detection \cite{besse2018}.   The ability to realize low noise, quantum limited microwave photon detectors, and dc-SQUID arrays comprising thousands of unit cells operating at a few tens of mK temperatures and below\cite{macklin2015}, shows promise for demonstrating strong microwave Hawking radiation signals.

Looking forward, much remains to be explored; the JTWPA design \cite{macklin2015} needs to be adapted so as to enable propagating flux pulse/step biasing \cite{zorin2019}, and  microwave photon detection circuitry suitable for verifying correlated photon pairs needs to be developed.

In terms of theory modeling, an analysis is required of the photon production from vacuum starting with the discrete Lagrangian (\ref{squidarrayeq}). This will allow us to determine the effect of the  `Planck' scale physics set by the unit cell length $a$  on the Hawking radiation prediction following from the approximate continuum model \cite{unruh1981,schutzhold2005}. For such an analysis, we can assume in the first instance that the phase field fluctuations are small in magnitude, hence harmonically approximating the nonlinear cosine potential in Eq. (\ref{squidarrayeq}). 

An interesting question to be addressed concerns the consequence of applying a non-zero, uniform (i.e. constant) external flux bias `floor' over the length of the dc-SQUID array \cite{chow1998}. This reduces the effective Josephson energy $E_J \cos\left(\pi\Phi_n^{\mathrm{ext}}/\Phi_0\right)$, somewhat analogous to weakening the spring constant of a harmonic oscillator mass. Zero-point fluctuations (i.e., quantum uncertainty) in the phase coordinate operators $\hat{\varphi}_n$ will correspondingly increase and the full, nonlinear cosine potential in the discrete Lagrangian (\ref{squidarrayeq}) may need to be taken into account. If we now add this constant flux bias floor to the propagating flux bias front, what will be the consequence for photon production from vacuum? In an analogous sense, increasing quantum uncertainty in the phase coordinate operators may be interpreted as increasing fluctuations in the effective metric and hence fluctuating event horizon location. The ability to tunably control such fluctuations through the externally applied flux is a unique feature of the present superconducting circuit analogue.

\section{Analogue Oscillatory Unruh Effect}
In the preceding section, we saw how Josephson junction elements enhance the functionality of superconducting microwave transmission lines, and in particular enable the realization of an effective event horizon and accompanying Hawking radiation. In this section, we consider a further enhancement in the functionality of superconducting circuits by bringing into play a special type of mechanical oscillator called a `film bulk acoustic resonator' (FBAR) \cite{cuffe2013,sanz2018}. Such crystal (or crystalline) mechanical oscillators may be fashioned out of silicon or other materials and undergo dilatational (i.e., breathing) mode oscillations in the few to tens of GHz frequency range depending on their thickness (typically a few hundred nm). These mechanical frequencies are large enough to match those of microwave cavities; with the upper and lower FBAR surfaces metalized and having areas of the order of a few hundreds of $\mu{\mathrm{m}}^2$, the dilational mechanical motion can strongly couple capacitively to microwave cavity fields. This then allows the possibility to realize close analogues of the dynamical Casimir effect for mechanically oscillating (i.e., accelerating) mirrors \cite{sanz2018}   and the Unruh effect for mechanically oscillating photodetectors \cite{doukas2013,wang2019}. By `Unruh effect', we adopt here a broader definition than is conventionally considered, where an arbitrarily accelerating photodetector registers emitted photons from the Minkowski vacuum, and for which their energy spectrum does not necessarily follow an effective thermal distribution.  Furthermore, there need not be an event horizon in the detector's centre of mass rest frame. The usual convention is to restrict to the limiting ideal case of an eternal, constant proper acceleration for which the detected photon energy spectrum coincides with a thermal distribution. In order to distinguish from the conventional Unruh effect, we shall use the term `oscillatory Unruh effect', reflecting the nature of the acceleration in the present considered scheme.   

\begin{figure}[!htb]
\centering\includegraphics[width=4.5in]{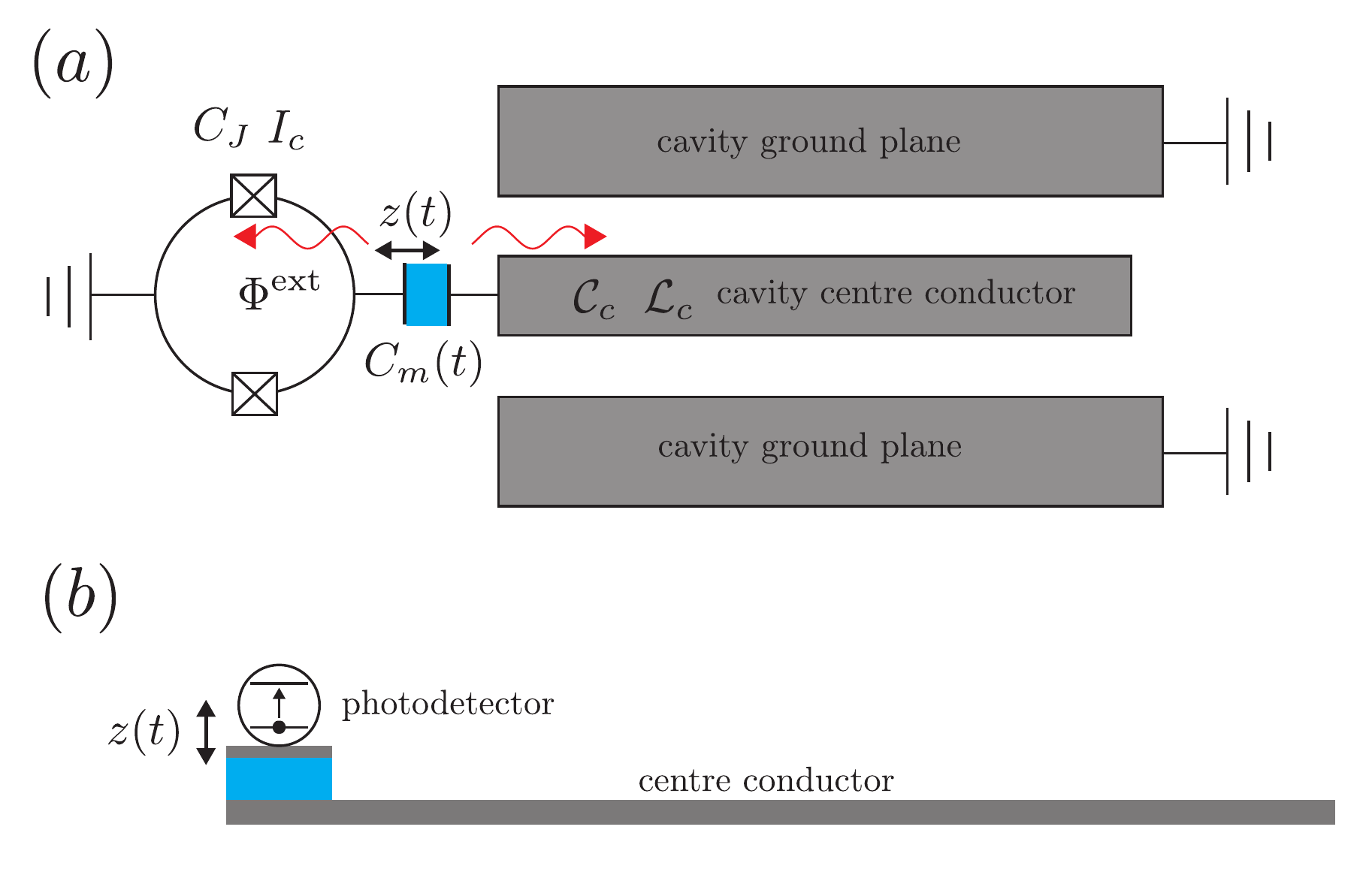}
\caption{(a) Circuit diagram of the analogue oscillatory Unruh effect scheme. The dc-SQUID serves as a superconducting qubit photodetector, which couples to a co-planar waveguide cavity via an FBAR capacitor (blue). (b) Side view of the scheme indicating the placement of the photodetector atop the FBAR. Not shown are the qubit-cavity probe circuitry and means of FBAR mechanical actuation.}
\label{unruhfig}
\end{figure}

We will now outline a possible scheme for realizing the oscillatory Unruh effect that utilizes a microwave cavity, a dc-SQUID, and an FBAR  (Fig. \ref{unruhfig}). The scheme is somewhat related to our recent proposal \cite{wang2019}, but with the key difference that the dc-SQUID functions as a qubit \cite{vool2017}  photodetector instead of using another coupled microwave cavity as photodetector. This enables the localized siting of the photodetector atop the accelerating FBAR (Fig. \ref{unruhfig}b), in contrast to our previous scheme \cite{wang2019} where the detector had only a small geometrical overlap with the cavity via their mutual coupling capacitance. When the FBAR is mechanically driven at say its fundamental transverse breathing mode frequency $\omega_m$, then the qubit undergoes actual oscillatory acceleration at this frequency, modulating its coupling to the microwave cavity via the FBAR capacitance $C_m(t)$ (where the `$m$' subscript denotes `mechanical'). Under the frequency matching condition $\omega_m=\omega_c+\Delta E_{qb}/\hbar$, where $\omega_c$ is the fundamental mode frequency of the cavity and $\Delta E_{qb}$ is the qubit energy level spacing, we can have resonant enhancement of correlated microwave photon pair production from vacuum for realizable large, superconducting cavity quality factors, with one photon in the pair appearing in the cavity and the other photon absorbed by the detector, inducing a transition to its excited state. By utilizing non-piezoelectric materials such as silicon for the  FBAR, oscillatory dipole radiation due to induced capacitor plate surface charges is minimized; the photon pair production from vacuum may then be viewed as a consequence of the photodetector's actual acceleration. Hence the present scheme may be viewed as a close analogue of the actual oscillatory Unruh effect.       

Applying Kirchhoff's laws to the circuit shown in Fig. \ref{unruhfig}a, we obtain the following dynamical equations in terms of the dc-SQUID average phase coordinate $\varphi = \left(\varphi_1 + \varphi_2\right)/2$, (with $\varphi_1$, $\varphi_2$ the gauge invariant phases across each JJ), the phase coordinate $\varphi_m$ across the FBAR capacitance $C_m(t)$, and the cavity phase field coordinate $\varphi_c(x,t)$ between the centre conductor $x$ location and ground:
\begin{equation}
2C_J \frac{\Phi_0} {2\pi} \Ddot{\varphi} + 2I_c\cos \left( \frac{\pi\Phi^{\mathrm{ext}}} {\Phi_0} \right)\sin\varphi - \frac{\Phi_0} {2\pi} \frac{d}{dt} \left(C_m(t) \frac{d\varphi_m}{dt} \right) = 0,
\label{kirch7eq}
\end{equation}
\begin{equation}
\frac{\partial^2 \varphi_c} {\partial x^2} - {\mathcal{L}_c\mathcal{C}_c}\frac{\partial^2 \varphi_c} {\partial t^2} = 0,
 \label{kirch9eq}
\end{equation}
with the boundary conditions
\begin{equation}
 \frac{\Phi_0} {2\pi \mathcal{L}_c} \frac{\partial \varphi_c(0,t)} {\partial x}=\frac{\Phi_0} {2\pi} \frac{d}{dt} \left(C_m(t) \frac{d\varphi_m}{dt} \right)
\label{kirch8eq}
\end{equation}
and
\begin{equation}
\frac{\partial \varphi_c(L,t)}{\partial x} = 0.
\label{boundaryco}
\end{equation}
The latter condition follows from the vanishing of the current at the right, terminated end of the center conductor of length $L$. The FBAR capacitance satisfies
\begin{equation}
C_m(t) = C_m^{(0)}(1+z(t)/D),
\label{fbarcap}
\end{equation}
where $z(t) = A \cos(\omega_mt)$ is the driven dilatational displacement of the FBAR, with $A$ the amplitude, and $D$ the FBAR thickness; we assume that $|z(t)|\ll D$. The FBAR capacitance phase field coordinate can be eliminated through the following phase coordinate relation that follows from the voltage Kirchhoff law:
\begin{equation}
\varphi_m(t)=\varphi_c(0,t)-\varphi(t).
\label{kirch10eq}
\end{equation}

We approximate the cavity phase field $\varphi_c (x,t)$ as a series expansion in terms of eigenfunctions $\phi_n(x)$:
\begin{equation}
 \varphi_c (x,t) = \sum_n q_n(t)\phi_n(x).
 \label{modedeceq}
\end{equation}
Assuming that $C_m^{(0)}\ll{\mathcal{C}}_c L$, we
replace Eq. (\ref{kirch8eq}) with the approximate boundary condition $\varphi_c^{\prime} (0,t) \approx 0$, and taking into account the opposite end boundary condition (\ref{boundaryco}), the orthogonal eigenfunctions are
\begin{equation}
\phi_n(x) = \cos\left( \frac{n\pi x} {L}\right),\, n=1,2,\dots
\label{modeseq}
\end{equation}
From Eqs. (\ref{modedeceq}) and (\ref{modeseq}), the to be determined cavity phase mode coordinates $q_n(t)$ are given as
\begin{equation}
q_n(t) = \frac{2}{L} \int_0^L dx \varphi_c(x,t) \phi_n(x).
\label{coefficient1}
\end{equation}
Differentiating (\ref{coefficient1}) twice with respect to time and applying the cavity wave equation (\ref{kirch9eq}), we have
\begin{equation}
\Ddot{q}_n(t) = \frac{2}{\mathcal{L}_c \mathcal{C}_c L} \int_0^L dx \varphi_c^{\prime\prime} (x,t) \phi_n(x).
\label{coefficient2}
\end{equation}
Integrating (\ref{coefficient2}) by parts twice, applying the eigenvalue equation $\phi_n^{\prime \prime}(x) = -k_n^2 \phi_n$ (with $k_n=n\pi/L$) and the boundary conditions  (\ref{kirch8eq}) and (\ref{boundaryco}), we obtain
\begin{equation}
\Ddot{q}_n(t) = - \frac{2} {\mathcal{C}_c L} \frac{d}{dt} \left[C_m(t) \left(\dot{\varphi_c} (0,t)  - \dot{\varphi} \right) \right] - \frac{k_n^2} {\mathcal{L}_c \mathcal{C}_c} q_n(t).
\label{coefficient3}
\end{equation}
When the resonance condition $\omega_m = \omega_n + \Delta E_{qb}/\hbar$ is satisfied for some cavity mode frequency $\omega_{n}=k_n^2 /\mathcal{L}_c \mathcal{C}_c$ and qubit level spacing $\Delta E_{qb}$ (see below),  we can apply the single-mode approximation to the cavity. Equation (\ref{coefficient3}) then becomes
\begin{equation}
\Ddot{q}_c + \frac{2} {\mathcal{C}_c L} \frac{d}{dt} \left[C_m(t) \left(\dot{q_c}  - \dot{\varphi} \right) \right] + \omega_c^2 q_c = 0,
\label{coefficient4}
\end{equation}
where $\omega_c^2 = \omega_n^2 /\mathcal{L}_c \mathcal{C}_c$, $q_c(t) = q_n(t)$ and we have used $\phi_n(0)=1$.

Using Eqs. (\ref{kirch10eq}) and (\ref{fbarcap}), Eq. (\ref{kirch7eq}) becomes
\begin{equation}
\frac{\Phi_0} {2\pi} 2C_J \Ddot{\varphi} - \frac{\Phi_0} {2\pi} \frac{d}{dt} \left[C_m{(t)} (\dot{q_c} - \dot{\varphi}) \right] + 2I_c\cos \left( \frac{\pi\Phi^{\mathrm{ext}}} {\Phi_0} \right) \sin\varphi = 0.
\label{varphieq}
\end{equation}
The coupled cavity and dc-SQUID equations of motion (\ref{coefficient4}) and (\ref{varphieq}) follow via the Euler-Lagrange equations from the following Lagrangian:
\begin{equation}
L = \left(\frac{\Phi_0} {2\pi}\right)^2 \left[ \frac{C_c + C_m{(t)}}{2} \dot{q}_c^2 - \frac{C_c \omega_c^2}{2} q_c^2 + \frac{2C_J + C_m{(t)}} {2} \dot{\varphi}^2 - C_m{(t)} \dot{q_c} \dot{\varphi} \right] + E_J \cos\varphi,
\label{lagrangeeq}
\end{equation}
where $C_c=\mathcal{C}_c L/2$ is the cavity mode lumped element capacitance and $E_J = \frac{I_c \Phi_0}{\pi} \cos \left( \frac{\pi\Phi^{\mathrm{ext}}} {\Phi_0} \right)$ is the effective Josephson energy
which can be tuned by the external flux $\Phi^{\mathrm{ext}}$.

Performing the Legendre transformation on the Lagrangian, we obtain the following
cavity dc-SQUID system  Hamiltonian:
\begin{eqnarray}
H&=&\left(\frac{2\pi} {\Phi_0}\right)^2 \frac{p_c^2}{2C_c} + \left(\frac{\Phi_0} {2\pi} \right)^2\frac{C_c \omega_c^2 q_c^2} {2} + \frac{E_C}{\hbar^2} p_{\varphi}^2 - E_J \cos\varphi \cr
&&+ \left(\frac{2\pi} {\Phi_0}\right)^2\frac{C_m^{(0)}} {C_c C_{\Sigma}} \left(1 + \frac{2C_J} {C_{\Sigma}} \frac{z(t)}{D} \right) p_c p_{\varphi},
\label{hamiltoneq}
\end{eqnarray}
where $E_C=(2e)^2/(2 C_{\Sigma})$, $C_{\Sigma}=2C_J+C_m^{(0)}$, is the Cooper-pair charging energy of the dc-SQUID island formed by the two JJ's and FBAR capacitor, and $p_c$ and $p_{\varphi}$ are the momentum coordinates canonically conjugate to the $q_c$ and $q_{\varphi}$ coordinates, respectively. Expression (\ref{hamiltoneq}) for the Hamiltonian assumes that $C_m^{(0)}\ll C_c$.

Quantizing, Hamiltonian (\ref{hamiltoneq}) can be expressed as the following corresponding operator:
\begin{equation}
    \hat{H}=\hbar\omega_c \hat{a}^{\dag}\hat{a} +E_C \hat{n}^2 -E_J \cos\hat{\varphi}-i\sqrt{\frac{R_K}{4\pi Z_c}} E_C \frac{C_m^{(0)}} {C_c} \left(1 + \frac{2C_J} {C_{\Sigma}} \frac{z(t)}{D} \right) \left(\hat{a}-\hat{a}^{\dag}\right)\hat{n},
    \label{qhamiltoneq}
\end{equation}
where $\hat{n}=\hat{p}_{\varphi}/\hbar$ is the number operator for Cooper-pairs transferred to the dc-SQUID island, $Z_c=1/(\omega_c C_c)$ is the cavity mode impedance, and $R_K=h/e^2\approx 25.8~{\mathrm{K}}\Omega$ is the von Klitzing resistance. In order to realize Hamiltonian (\ref{qhamiltoneq}), the dc-SQUID circuit shown in Fig. \ref{unruhfig} should in practise also include a gate voltage bias in order to be able to cancel out any stray charge on the dc-SQUID island.
Expressing the Hamiltonian operator (\ref{qhamiltoneq}) in terms of the decoupled dc-SQUID Hamiltonian energy eigenstates $|i\rangle$ and eigenvalues $E_i$, $i=0,1,2,\dots$, we have in the truncated, two lowest dc-SQUID energy eigenstate approximation:
\begin{equation}
    \hat{H}=\hbar\omega_c \hat{a}^{\dag}\hat{a} +\frac{\Delta E_{qb}}{2} \sigma_z-i\hbar \left(g\sigma^+ +g^*\sigma^-\right) \left(\hat{a}-\hat{a}^{\dag}\right) -i\hbar\cos\left(\omega_m t\right) \left(g_m\sigma^+ +g_m^*\sigma^-\right) \left(\hat{a}-\hat{a}^{\dag}\right),
    \label{approxhameq}
\end{equation}
where $\Delta E_{qb}=E_1-E_0$,  the cavity-qubit coupling strength is 
\begin{equation}
    g=\hbar^{-1}\sqrt{\frac{R_K}{4\pi Z_c}} E_C \frac{C_m^{(0)}} {C_c} \langle 1|\hat{n}|0\rangle,
    \label{geq}
\end{equation}
the driven, FBAR dispacement cavity-qubit coupling strength is 
\begin{equation}
   g_m=\frac{2 C_J}{C_{\Sigma}}\frac{A}{D} g,
    \label{lambdaeq}
\end{equation}
and we have assumed the resonance condition $\hbar\omega_m=\hbar\omega_c +\Delta E_{qb}$, required for the validity of the above two-state truncation.  
We furthermore require that 
$\omega_c\neq\Delta E_{qb}/\hbar$; in particular, the cavity and qubit frequencies must differ by more than their respective relaxation rates (see below). The latter condition is required so that photon pairs are not generated directly in the cavity when $\omega_m=2\omega_c$, which would correspond to the dynamical Casimir effect; in deriving the Hamiltonian (\ref{hamiltoneq}) above, we neglected direct mechanical modulation of the cavity by dropping the $C_m(t)$ contribution to the cavity kinetic energy term. 

Hamiltonian (\ref{approxhameq}) coincides with the familiar Rabi Hamiltonian, but with modulated Rabi coupling $g+g_m \cos(\omega_m t)$ \cite{liberato2009}. With the resonance condition $\hbar\omega_m=\hbar\omega_c +\Delta E_{qb}$, we expect correlated production of a photon in the cavity mode vacuum and the absorption of a photon by the qubit detector (i.e., excitation of the qubit from its ground state) due to the mechanically driven FBAR capacitance that couples the cavity and detector. In the presence of unavoidable cavity and qubit damping, given by some rates $\gamma_b$ and $\gamma_{qb}$, respectively, the average cavity photon number and the photon production rate will reach a steady state for times that are long compared to $\gamma_b^{-1}$, $\gamma_{qb}^{-1}$, which we will now estimate in the following.

As example, realistic parameters, we consider a silicon FBAR with fundamental dilatational mode frequency $\omega_m\approx 2\pi\times 10~{\mathrm{GHz}}$, corresponding to a thickness $D\approx 500~{\mathrm{nm}}$ and capacitance $C_m^{(0)}\sim 10~{\mathrm{fF}}$ assuming a $100 ~\mu{\mathrm{m}}^2$ FBAR capacitor plate area \cite{wang2019}. With a co-planar cavity characteristic impedance $Z_c\approx 50~\Omega$ and for $\omega_c\approx 2\pi\times 5~{\mathrm{GHz}}$, we have for the cavity lumped element capacitance $C_c\approx 1~{\mathrm{pF}}$. 
Assuming $C_J\approx C_m^{(0)}$, we have approximately for the dc-SQUID charging energy $E_C/h\approx 1~{\mathrm{GHz}}$. For a Josephson energy $E_J/h\sim 15~{\mathrm{GHz}}$, we obtain $\Delta E_{qb}/h\approx 5~{\mathrm{GHz}}$, ensuring that the resonance condition $\omega_m=\omega_c+\Delta E_{qb}/\hbar$ can be satisfied; fine tuning of $\Delta E_{qb}$ can be achieved by varying the external flux bias $\Phi^{\mathrm{ext}}$. With a Josephson to Cooper-pair charging energy ratio $E_J/E_C\approx 15$, the qubit is of the so-called `transmon' type \cite{koch2007}. With these parameters, the cavity-qubit coupling strength (\ref{geq}) is $|g|\approx 2\pi\times 0.5~{\mathrm{GHz}}$. In order to avoid producing photon pairs directly in the cavity (dynamical Casimir effect), we require that $|\Delta E_{qb}/\hbar-\omega_c|\gg \gamma_c,\, \gamma_{qb}$, as mentioned above. For achievable FBAR dilatational amplitude $A\approx 10^{-11}~{\mathrm{m}}$ \cite{cuffe2013},  the FBAR displacement cavity-qubit coupling (\ref{lambdaeq}) then becomes $|g_m|\approx 10^5 ~\mathrm{s}^{-1}$. 

\begin{figure*}[thb]
\centering
\subfloat{\label{avephotonfig}
  \includegraphics[width=0.45\textwidth]{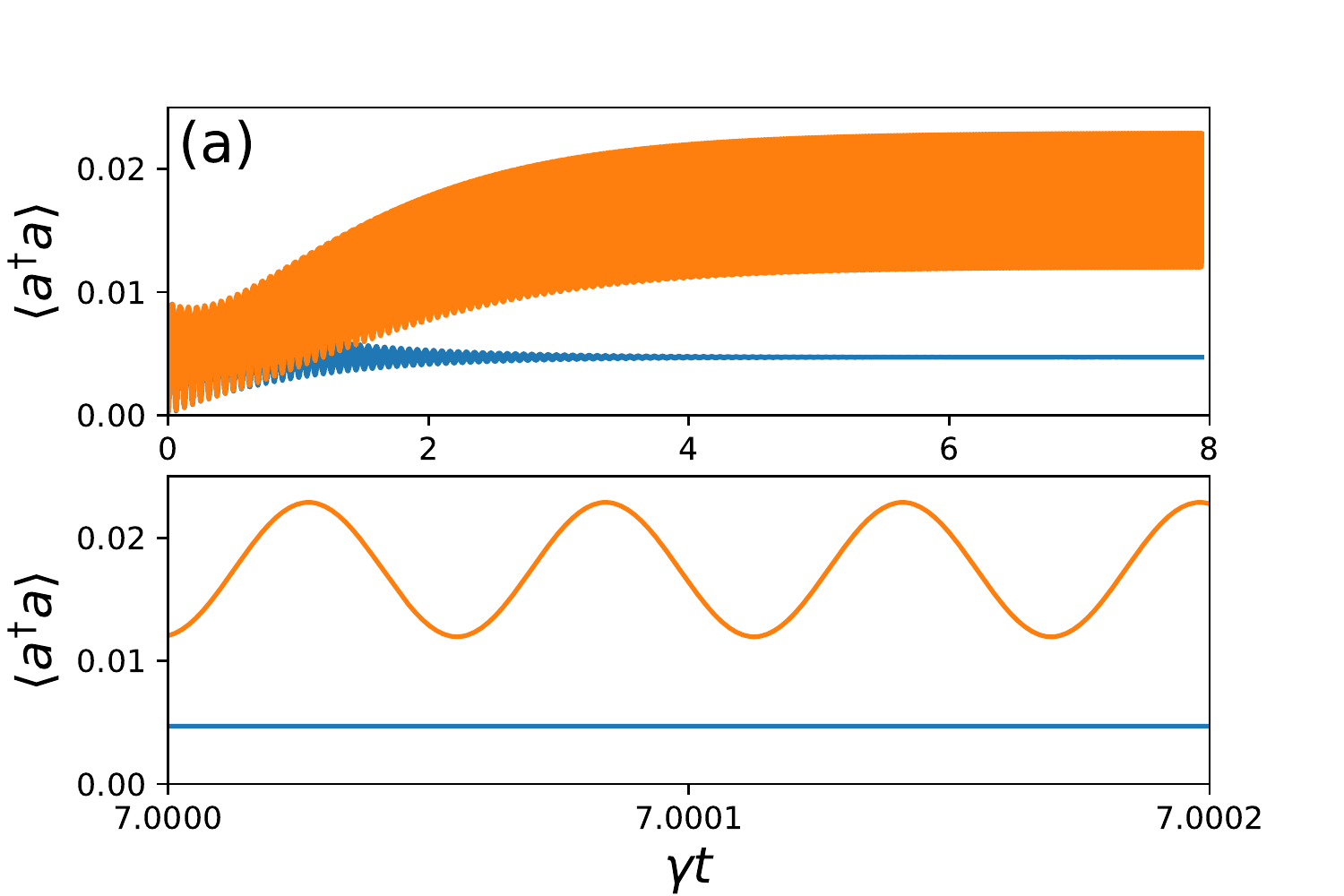}}
\subfloat{\label{lognegfig}
  \includegraphics[width=0.45\textwidth]{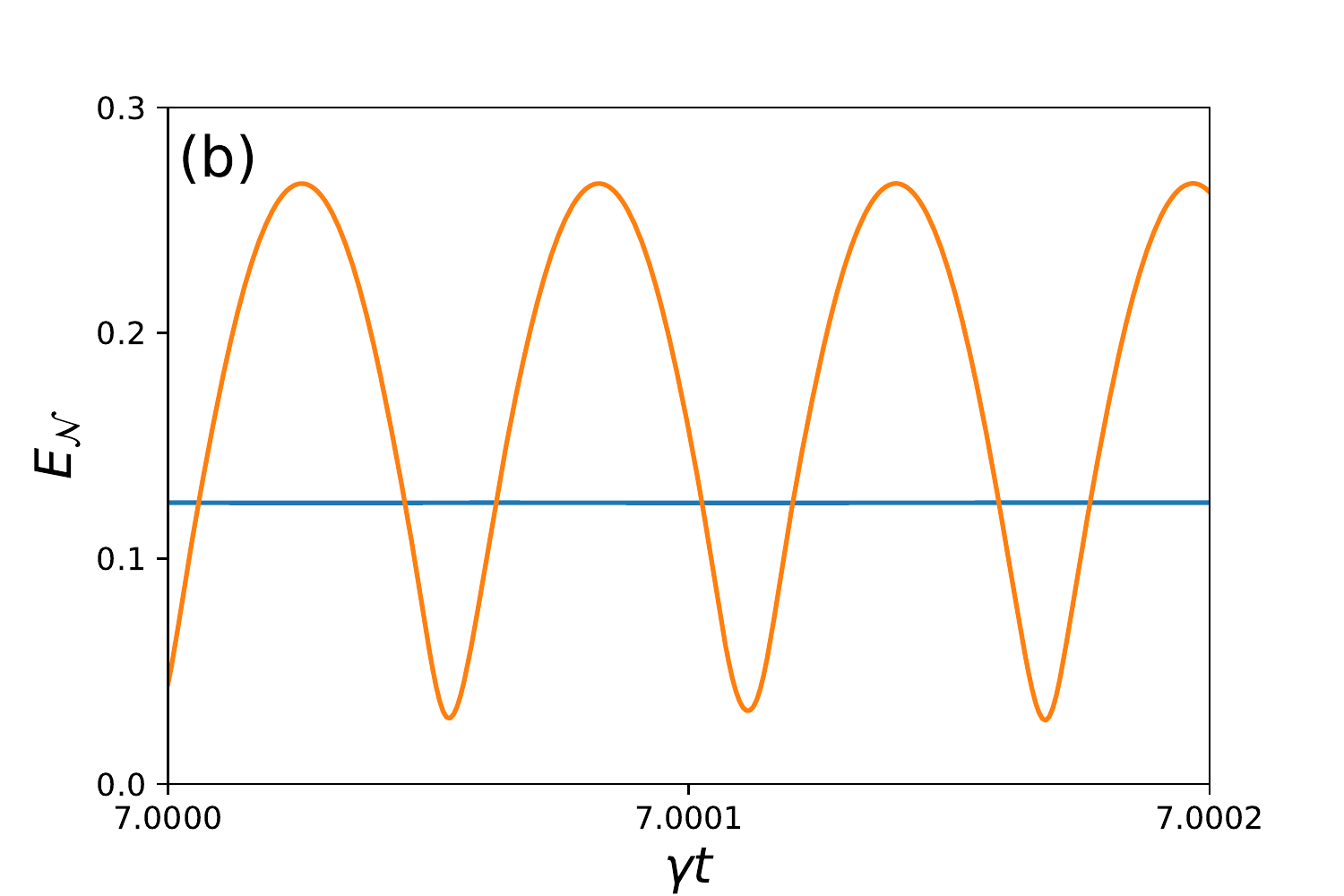}
  }
\caption{Photon production from cavity vacuum. (a) The upper panel shows the evolution of the average photon number $\langle \hat{a}^{\dag}\hat{a}\rangle$ for the cavity system mode ranging from the initial bare cavity ground state to the long-time limit steady state; the lower panel shows the zoomed-in average cavity photon number in the long-time limit. (b) Evolution of the logarithmic negativity entanglement measure $E_{\mathcal{N}}$ in the long-time limit. The orange lines are for the renormalized, resonant case where $\omega_m = 2.202791\omega_c$ and blue lines for the off-resonance case; note that the resonance condition gets renormalized because of the non-negligible $g$ value.  The other parameter values are $\Delta E_{qb}/\hbar = 1.1\omega_c$, $g=0.1\omega_c$, $g_m=10^{-6}\omega_c$, and $\gamma_c = \gamma_{qb} = \gamma = 5\times 10^{-6} \omega_c$.}
\end{figure*} 

Assuming cavity and qubit damping rate magnitudes $\gamma_c\approx\gamma_{qb}\approx 10^{5}~{\mathrm{s}}^{-1}$, we obtain an average cavity photon number magnitude $\langle \hat{a}^{\dag}\hat{a}\rangle\approx 0.01$-$0.02$ in the long-time limit, steady state (Fig.\ref{avephotonfig}); the latter average is obtained by solving numerically the Lindblad master equation for Hamiltonian (\ref{approxhameq}) \cite{johansson2013}. A numerical solution is preferred over analytical approximation because the above derived, static Rabi coupling parameter magnitude is relatively strong ($g/\omega_c\approx 0.1$), resulting in non-negligible hybridization between the bare cavity and qubit energy eigenstates. As a consequence, the off-resonance average photon number is non-zero, while the  earlier stated resonance condition on $\omega_m$  for maximum correlated photon pair production involving the bare frequencies  (i.e., $\omega_m=\omega_c+\Delta E_{qb}/\hbar$) becomes renormalized. Given that $\langle \hat{a}^{\dag}\hat{a}\rangle\ll 1$ and using a detailed balance based estimate,  the cavity photon production rate in the steady state is $\gamma_c\langle \hat{a}^{\dag}\hat{a}\rangle\approx 10^3\, {\mathrm{s}}^{-1}$; such a rate should be measurable with current quantum limited microwave photodetection methods \cite{lahteen2016,sandbo2018,besse2018}. Note that the cavity subsystem reduced state in the long-time limit is not a thermal state, as evidenced in the numerically obtained non-zero $|\langle \hat{a}^2\rangle|\approx 0.06$. The non-thermality is a consequence of the relatively large, static Rabi coupling $|g|$ that results in cavity-qubit hybridization; in particular, for smaller $|g|$ coupling, we would expect the cavity reduced state to  be approximately thermal.

In Fig. \ref{lognegfig}, we plot the logarithmic negativity entanglement measure \cite{vidal2002} in the long time limit. The observed on-resonance entanglement peaks exceed the off-resonance entanglement (with the latter a consequence of  cavity-qubit hybridization mentioned above) periodically with the mechanical drive, and in phase with the average photon number maxima indicated in Fig. \ref{avephotonfig}. Such logarithmic negativity peaks signify quantum correlated photon pair production from vacuum.

Looking forward, a more complete analysis of the above, oscillatory Unruh effect scheme is required, in particular including the cavity mode and qubit state measurement circuitry, as well as the method of FBAR actuation. This will give a better idea of the feasibility of the scheme, including how to verify correlated photon production from vacuum, as opposed to typically unavoidable microwave radiation noise. It will also be interesting to consider a semi-infinite transmission line \cite{sanz2018} instead of a finite length microwave cavity, such that the microwave vacuum to which the accelerating qubit photodetector couples comprises a dense spectrum of propagating modes. Such a set-up will facilitate a clearer picture of how close is the present scheme to an actual realization of an oscillatory, accelerating photodetector registering photons from the Minkowski vacuum \cite{doukas2013}. 

In the above treatment, the FBAR was treated as a classical, driven mechanical oscillator. However, given its micron-scale size and GHz range fundamental oscillation frequency, such FBARs can also display quantum behaviour \cite{oconnell2010}. It would be interesting to consider the consequences for the oscillatory Unruh effect when the FBAR is in a quantum state, such as a superposition of two different dilatational amplitudes $A$, hence furnishing a close analogue of a quantum non inertial reference frame for the qubit photodetector \cite{katz2015}.   

\section{Conclusion}
In the present work, we have described possible schemes for realizing analogues of the Hawking and (oscillatory) Unruh effects utilizing superconducting microwave circuits with Josephson junction and film bulk acoustic resonator elements. Superconducting circuits are in principle ideally suited for demonstrating such microwave photon production from vacuum processes, due to their low noise, controllability and existing advanced microwave fabrication and quantum-limited detector capabilities. We also mapped out some possible directions for near future investigation; it is our view that  analogue gravity phenomena involving superconducting circuits is a rich area that has yet to be comprehensively explored.   

\aucontribute{MPB conceived the project and carried out the analogue Hawking effect analysis. HW carried out the analogue Unruh effect analysis. Both authors drafted, read and approved the manuscript.}

\competing{The authors declare that they have no competing interests.}

\funding{ This work was supported by the NSF under Grant No. DMR-1507383}

\ack{We thank the organizers of the  `The next generation of analogue gravity experiments' Royal Society workshop for providing a stimulating venue in which to present this work. We also thank A. Kamal, P. D. Nation, W. D. Oliver, A. J. Rimberg, and K. Sinha for helpful and inspiring conversations.}


\begin{thebibliography}{9}

\bibitem{nation2009} Nation PD, Blencowe MP, Rimberg AJ, Buks E. 2009 Analogue Hawking radiation in a dc-SQUID array transmission line. \textit{Phys. Rev. Lett.} \textbf{103}, 087004.

\bibitem{schutzhold2005} Sch\"{u}tzhold R, Unruh WG. 2005 Hawking radiation in an electromagnetic waveguide? \textit{Phys. Rev. Lett.} \textbf{95}, 031301.

\bibitem{philbin2008} Philbin TG {\textit et al}. 2008 Fibre-optical analog of the event horizon. \textit{Science} \textbf{319}, 1367.


\bibitem{wilson2011} Wilson CM \textit{et al}. 2011 Observation of the dynamical Casimir effect in a superconducting circuit.  \textit{Nature (London)}
\textbf{47}, 376.

\bibitem{lahteen2013} L\"{a}hteenm\"{a}ki P, Paraoanu GS, Hassel J, Hakonen PJ. 2013 Dynamical Casimir effect in a Josephson metamaterial.
\textit{Proc. Natl. Acad. Sci. USA}  \textbf{110}, 4234.

\bibitem{nation2012} Nation PD, Johansson JR, Blencowe MP, Nori F. 2012 Stimulating uncertainty: Amplifying the quantum vacuum with superconducting circuits. \textit{Rev. Mod. Phys.} \textbf{84}, 1.

\bibitem{sanz2018} Sanz M, Wieczorek W, Gr\"{o}blacher S, Solano E. 2018 Electro-mechanical Casimir effect. \textit{Quantum} \textbf{2}, 91.

\bibitem{wang2019} Wang H, Blencowe MP, Wilson CM, Rimberg AJ. 2019 Mechanically generating entangled photons from the vacuum: A microwave circuit-acoustic analog of the oscillatory Unruh effect. \textit{Phys. Rev. A} \textbf{99}, 053833.

\bibitem{tian2017} Tian Z, Jing J, Dragan A. 2017 Analog cosmological particle generation in a superconducting circuit. \textit{Phys. Rev. D} \textbf{95}, 125003.

\bibitem{lang2019} Lang S, Sch\"{u}tzhold R. (2019) Analog of cosmological particle creation in electromagnetic waveguides \textit{Phys. Rev. D} \textbf{100}, 065003. 

\bibitem{macklin2015} Macklin C {\it et al}. 2015 A near-quantum-limited Josephson traveling-wave parametric amplifier. \textit{Science} \textbf{350}, 307.

\bibitem{vool2017} Vool U, Devoret M. 2017 Introduction to quantum electromagnetic circuits. \textit{Int. J. Circuit Theory Appl.} \textbf{45}, 897.

\bibitem{zorin2019} Zorin AB. 2019 Flux-driven Josephson traveling-wave parametric amplifier. \textit{Phys. Rev. Appl.} \textbf{12}, 044051.

\bibitem{unruh2003} Unruh WG, Sch\"{u}tzhold R. 2003 On slow light as a black hole analogue. \textit{Phys. Rev. D} \textbf{68}, 024008. 

\bibitem{nova2019} de Nova JRM, Golubkov K, Kolobov VI, Steinhauer J. 2019 Observation of thermal Hawking radiation and its temperature in an analogue black hole, \textit{Nature} \textbf{569}, 688.

\bibitem{balbinot2008}Balbinot R \textit{et al}. 2008 Nonlocal density correlations as a signature of Hawking radiation from acoustic black holes, \textit{Phys. Rev. A} \textbf{78}, 021603(R).

\bibitem{vidal2002} Vidal G, Werner RF. 2002 Computable measure of entanglement \textit{Phys. Rev. A} {\bf 65}, 032314.

\bibitem{jacquet2020}Jacquet MJ, K\"{o}nig F. 2020 The influence of spacetime curvature on quantum emission in optical analogues to gravity, \textit{arXiv}:2001.05807. 

\bibitem{eichler2012} Eichler C, Bozyigit D,  Wallraff A. 2012 Characterizing quantum microwave radiation and its entanglement with superconducting qubits using linear detectors, \textit{Phys. Rev. A} \textbf{86}, 032106.

\bibitem{lahteen2016}L\"{a}hteenm\"{a}ki P, Paraoanu GS, Hassel J, Hakonen PJ. 2016 Coherence and multimode correlations from vacuum fluctuations in a microwave superconducting cavity, \textit{Nature Commun.} \textbf{7}, 12548.

\bibitem{sandbo2018} Sandbo Chang CW \textit{et al}. 2018 Generating multimode entangled microwaves with a superconducting parametric cavity, \textit{Phys. Rev. Appl.} \textbf{10}, 044019.

\bibitem{besse2018} Besse J-C \textit{et al}. 2018 Single-shot quantum nondemolition detection of individual itinerant microwave photons, \textit{Phys. Rev. X} \textbf{8}, 021003.

\bibitem{unruh1981} Unruh WG. 1981 Experimental black-hole evaporation? \textit{Phys. Rev. Lett}. \textbf{46}, 1351.


\bibitem{chow1998} Chow C, Delsing P,  Haviland DB. 1998 Length-scale dependence of the superconductor-to-insulator quantum phase transition in one dimension \textit{Phys. Rev. Lett}. \textbf{81}, 204.

\bibitem{koch2007} Koch J \textit{et al}. 2007 Charge-insensitive qubit design derived from the Cooper pair box \textit{Phys. Rev. A} \textbf{76}, 042319.


\bibitem{cuffe2013} Cuffe J \textit{et al}. 2013 Lifetimes of confined acoustic phonons in ultrathin silicon membranes \textit{Phys. Rev. Lett}. \textbf{110}, 095503.

\bibitem{liberato2009} De Liberato S \textit{et al}. 2009 Extracavity quantum vacuum radiation from a single qubit \textit{Phys. Rev. A} \textbf{80}, 053810.

\bibitem{johansson2013} Johansson JR, Nation PD, Nori F. 2013 QuTiP 2: A Python framework for the dynamics of open quantum systems \textit{Comput. Phys. Comm.} \textbf{184}, 1234.
 
\bibitem{doukas2013}Doukas J, Lin S-Y,  Hu BL,  Mann R. 2013 Unruh effect under non-equilibrium conditions: oscillatory motion of an Unruh-DeWitt detector \textit{J. High Energy
Phys.} \textbf{11}, 119.

\bibitem{oconnell2010}O'Connell AD \textit{et al}. 2010 Quantum ground state and single-phonon control of a mechanical resonator \textit{Nature} \textbf{464}, 697.

\bibitem{katz2015}Katz BN, Blencowe MP, Schwab KC. 2015 Mesoscopic mechanical resonators as quantum noninertial reference frames \textit{Phys. Rev. A} \textbf{92}, 042104.








\end{thebibliography}
\end{document}